\documentclass[
twocolumn,
reprint,
superscriptaddress,
 amsmath,amssymb,
 aps, physrev,
]{revtex4-2}

\usepackage{graphicx}
\usepackage{dcolumn}
\usepackage{bm}
\usepackage{csquotes}

\usepackage[caption=false]{subfig}
\usepackage{xcolor}
\usepackage{adjustbox}
\usepackage{hyperref}
\hypersetup{hypertex=true,
            colorlinks=true,
            linkcolor=blue,
            anchorcolor=blue,
            citecolor=blue}   
\usepackage{gensymb}

\usepackage{xparse,xcoffins}

\ExplSyntaxOn
\NewCoffin\imagecoffin
\NewCoffin\labelcoffin

\keys_define:nn { miguel/label }
 {
  label   .tl_set:N = \l_miguel_label_tl,
  labelbox .bool_set:N = \l_miguel_label_box_bool,
  labelbox .default:n = true,
  fontsize .tl_set:N = \l_miguel_label_size_tl,
  fontsize .initial:n = \normalsize,
  pos .choice:,
  pos/nw .code:n = \tl_set:Nn \l_miguel_label_pos_tl { left,up },
  pos/ne .code:n = \tl_set:Nn \l_miguel_label_pos_tl { right,up },
  pos/sw .code:n = \tl_set:Nn \l_miguel_label_pos_tl { left,down },
  pos/se .code:n = \tl_set:Nn \l_miguel_label_pos_tl { right,down },
  pos/n .code:n = \tl_set:Nn \l_miguel_label_pos_tl { hc,up },
  pos/w .code:n = \tl_set:Nn \l_miguel_label_pos_tl { left,vc },
  pos/s .code:n = \tl_set:Nn \l_miguel_label_pos_tl { hc,down },
  pos/e .code:n = \tl_set:Nn \l_miguel_label_pos_tl { right,vc },
  pos .initial:n = nw,
  unknown .code:n   = \clist_put_right:Nx \l_miguel_label_clist
                       { \l_keys_key_tl = \exp_not:n { #1 } }
 }
\clist_new:N \l_miguel_label_clist
\box_new:N \l_miguel_label_box
\box_new:N \l_miguel_label_image_box

\NewDocumentCommand{\labelimage}{O{}m}
 {
  \group_begin:
  \tl_clear:N \l_miguel_label_tl
  \clist_clear:N \l_miguel_label_clist
  \keys_set:nn { miguel/label } { #1 }
  \tl_if_empty:NTF \l_miguel_label_tl
   {
    #2
   }
   {
    \SetHorizontalCoffin\imagecoffin
     {
      #2
     }
    \SetHorizontalCoffin\labelcoffin
     {
      \raisebox{\depth}
       {
        \bool_if:NTF \l_miguel_label_box_bool
         { \fcolorbox{white}{white}{\l_miguel_label_size_tl\l_miguel_label_tl} }
         { \l_miguel_label_size_tl\l_miguel_label_tl }
       }
     }
    \SetVerticalPole\imagecoffin{left}{-0pt+\CoffinWidth\labelcoffin/2}
    \SetVerticalPole\imagecoffin{right}{\Width-3pt-\CoffinWidth\labelcoffin/2}
    \SetHorizontalPole\imagecoffin{up}{\Height+8pt-\CoffinHeight\labelcoffin/2}
    \SetHorizontalPole\imagecoffin{down}{3pt+\CoffinHeight\labelcoffin/2}
    \use:x{\JoinCoffins\imagecoffin[\l_miguel_label_pos_tl]\labelcoffin[vc,hc]} 
    \TypesetCoffin\imagecoffin
   }
   \group_end:
 }
\NewDocumentCommand{\setlabel}{m}
 {
  \keys_set:nn { miguel/label } { #1 }
 }

\ExplSyntaxOff

\begin{document}

\preprint{APS/123-QED}

\title{\textbf{The spin-transfer inverse Faraday effect in Au/YIG thin film bilayers} 
}
\author{Víctor H. Ortiz}
\email{email: vortizhe@fiu.edu}
\affiliation{Department of Physics, The Ohio State University, Columbus, Ohio 43210, USA}
\affiliation{Department of Mechanical and Materials Engineering, Florida International University, Miami, Florida 33174, USA}
\author{Daniel Russell}
\author{Fengyuan Yang}
\author{Roland K. Kawakami}
\email{email: kawakami.15@osu.edu}
\affiliation{Department of Physics, The Ohio State University, Columbus, Ohio 43210, USA}
\date{\today}

\begin{abstract}
The inverse Faraday effect (IFE)—the generation of an effective magnetic field in a material upon irradiation with circularly polarized light—offers a pathway to ultrafast, non-thermal control of spin dynamics. In metallic/ferromagnetic-insulator (NM/FMI) bilayers, IFE-driven spin polarization in the metallic layer can be an alternative to generate magnon excitations in the magnetic insulator which otherwise would be inaccessible with photon energies below the insulator bandgap, but this avenue has remained mostly unexplored. In this work, we investigate the regime of spin-transfer via IFE in Au/YIG bilayers, by means of ultrafast time-resolved magneto-optical experiments. Understanding the mechanism of this spin-transfer IFE in these NM/FMI heterostructures is crucial for several reasons: it elucidates the fundamental coupling between light’s angular momentum and collective spin modes at buried interfaces, it enables all-optical generation and manipulation of pure spin currents, and it paves the way toward THz-frequency spintronic devices leveraging the insulator’s long magnon lifetimes. 
\end{abstract}

\maketitle

\section{\label{sec:Intro}Introduction}

The inverse Faraday effect (IFE) is a fundamental nonlinear magneto-optical effect where a circularly polarized light transfers photon angular momentum into a material’s electronic system, generating a quasi-static magnetization. First predicted by Pitaevskii~\cite{pitaevskii1961electric} and later confirmed experimentally~\cite{VanDerZiel1965}, IFE affects the orbital motion of electrons, causing the spins of electrons to change due to spin-orbit interactions~\cite{Hertel2006}. Recent ab initio studies~\cite{Berritta2016,Scheid2019,Mishra2023} have parsed the separate spin and orbital contributions to IFE, revealing strong dependencies on both material composition and excitation frequencies. Multiple experimental studies have been performed in non-magnetic~\cite{Wilks2004, Kruglyak2005, Ortiz2023} and magnetic~\cite{Choi2017} metals, with IFE-driven magnetization dynamics being extended to femtosecond control in ferromagnetic metallic films~\cite{John2017, Kimel2019} and orthoferrites~\cite{Kimel2005}. 

Among magnetic insulators, rare-earth iron garnets are of great interest for the spintronics community because of their high Curie temperature, chemical stability and tunability of their magnetic properties by either strain engineering~\cite{Kubota2012,Ortiz2018,Rosenberg2018,OrtizEuIG2021,Warren2022} or stoichiometry modifications~\cite{Dumont2007,Banos-Lopez2018,Chen2025}, with yttrium iron garnet (YIG) being the most used due to its ultra-low magnetic damping~\cite{Du2013}. However, harnessing these properties in optospintronic technologies remains a significant challenge. The direct all-optical, non-thermal control of YIG's magnetization is inherently inefficient and often requires either disruptive methods (such as material doping~\cite{Atoneche2010,Stupakiewicz2017} or high-power excitations~\cite{Stupakiewicz2021}) or require micrometer thicknesses~\cite{Hansteen2006,Mikhaylovskiy2012}, which can hinder practical nanoscale applications.

An alternative pathway is to control YIG indirectly by interfacing it with a material that offers strong, efficient light-matter coupling. This is where the IFE in metals becomes critical. Gold (Au) is a particularly compelling candidate for this task, as it supports strong plasmonic resonances well into the near-IR and exhibits a sizeable magneto-optical IFE response in both continuous films~\cite{Ortiz2023} and nanostructured geometries~\cite{Cheng2020, Koralewski2023, Gonzalez-Alcalde2024}. Integrating YIG with Au thus creates a powerful heterostructure that marries YIG's spin coherence with Au's efficient optomagnetic coupling. This interface provides a pathway for purely optical, non-thermal angular momentum transfer at the nanometer scale, enabling the excitation of YIG's spin dynamics without parasitic charge currents or the need for high-powered lasers.

In this work, we demonstrate magnetization control in YIG thin films via spin transfer from inverse Faraday effect–induced spins in adjacent Au films, providing an alternative route for manipulating magnetization in magnetic insulators. By exciting the bilayer with sub-bandgap photons ($\hbar \nu = 1.55$\,eV, YIG's $E_\textrm{gap} = 2.4$\,eV), we isolate the IFE contribution from thermal effects and achieve nanoscale optical control of magnetization. Our results establish a novel platform for infrared-driven, nonthermal magnetic manipulation in magnetic insulator/metal heterostructures for optospintronic applications.

\section{\label{sec:Methods}Methods}
\subsection{\label{sec:Samp_prep}Sample preparation and characterization}

We grew epitaxial YIG thin-film samples using the off-axis sputtering technique, with thicknesses of 10 nm and 15 nm. The films were deposited on (111)-oriented Gd$_{3}$Sc$_{2}$Ga$_{3}$O$_{12}$ (GSGG) single crystal substrates  using off-axis sputtering technique\cite{Yang2018}. The YIG films were grown at a substrate temperature of 650 $^\circ$C in a mixture of Ar + 1$\%$ O$_2$ gases with a total pressure of 11.5 mTorr. An RF power of 60 W was used for sputtering with a deposition rate of 20 nm/hour. Due to the lattice mismatch between YIG ($a_\textrm{YIG} = 12.376$\,\AA) and the GSGG substrate ($a_\textrm{GSGG} = 12.554$\,\AA), the tensile strain in the plane favors YIG magnetization to be oriented perpendicular to the sample plane. 

\begin{figure}[h]
\setlabel{pos=nw,fontsize=\normalsize} 
\subfloat{\labelimage[label=\textcolor{black}{(a)}]{\includegraphics[width=0.25\textwidth]{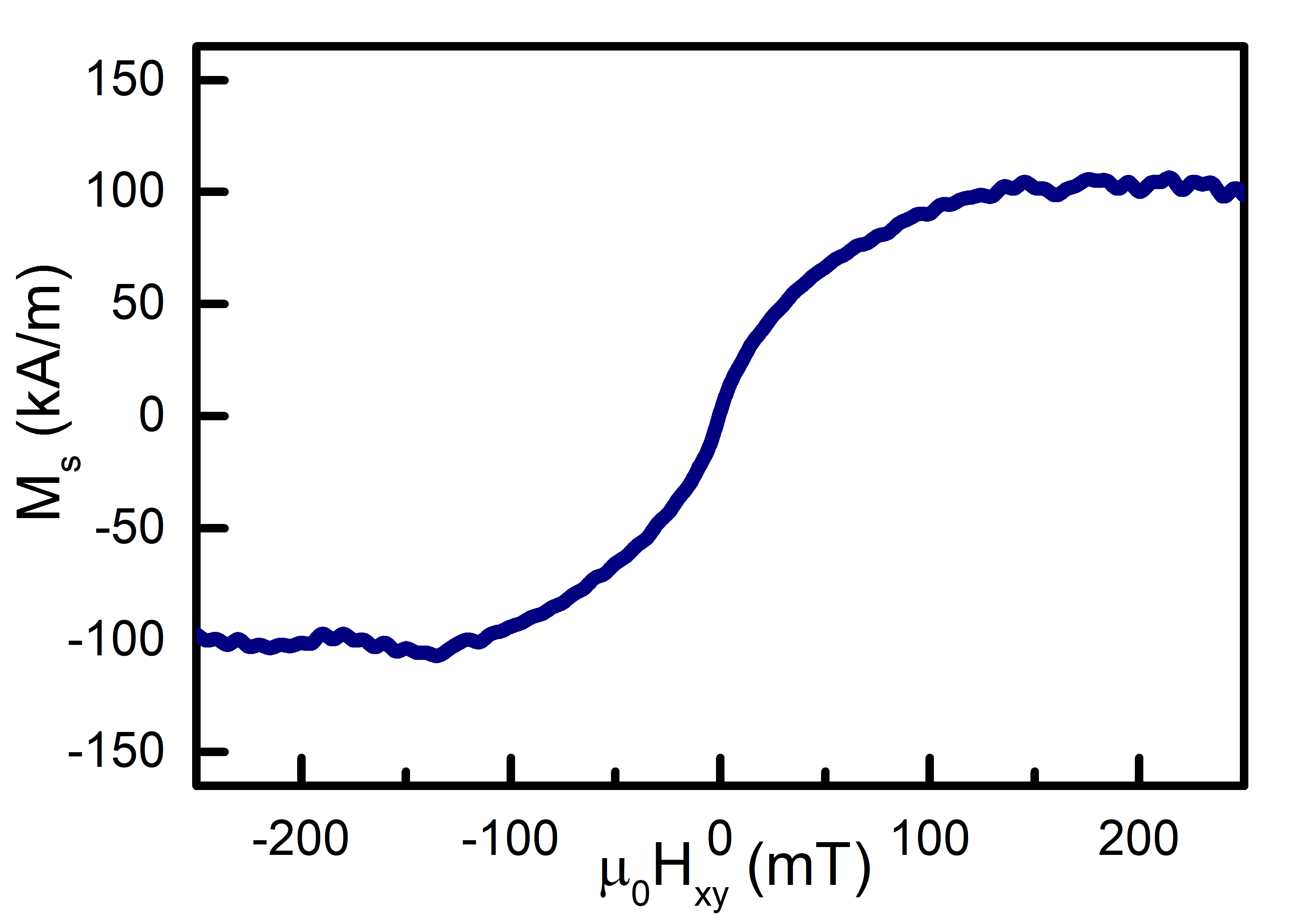}}\label{MvH10nm}}
\subfloat{\labelimage[label=\textcolor{black}{(b)}]{\includegraphics[width=0.25\textwidth]{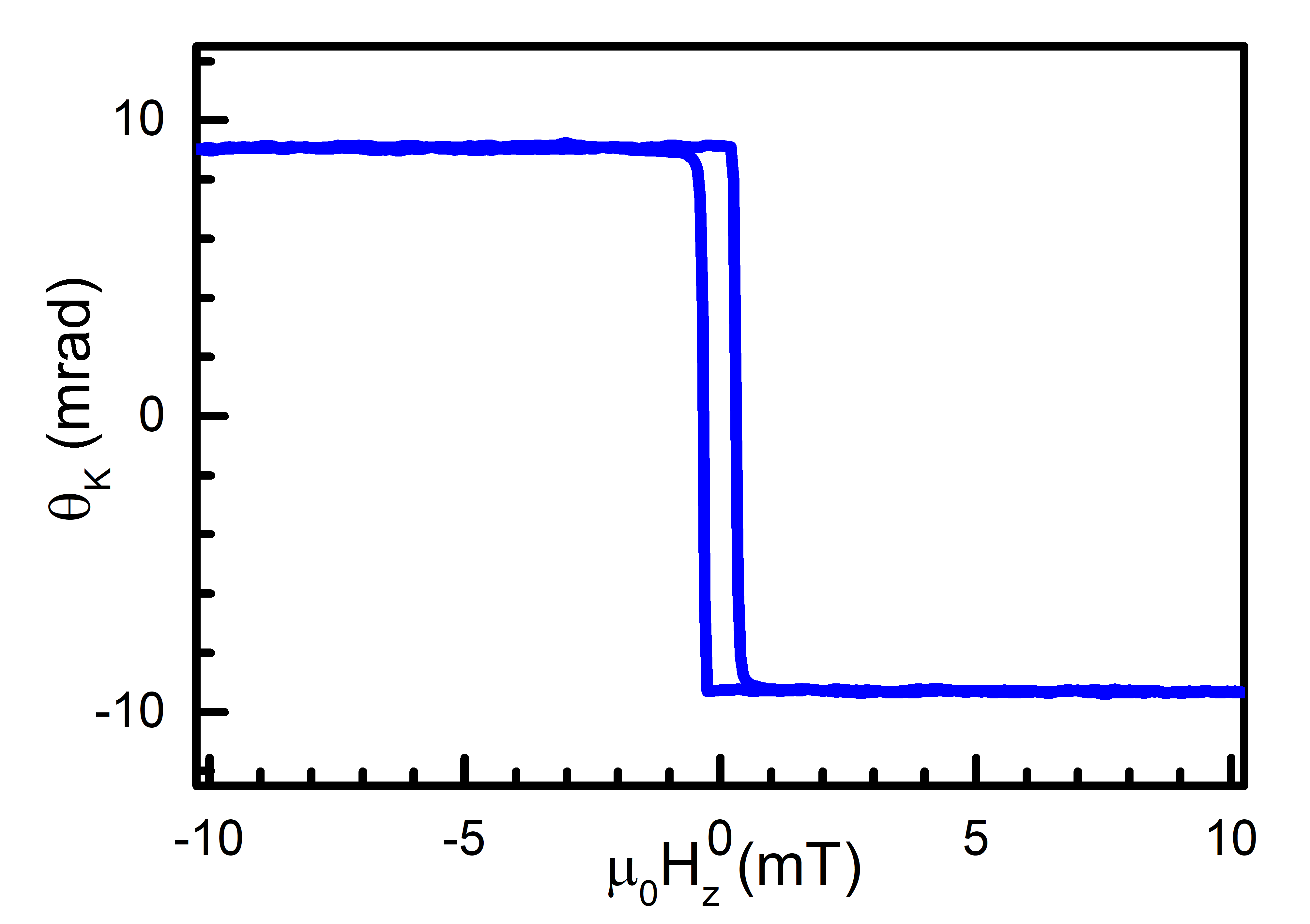}}\label{MOKE10nm}}
\hfill
\subfloat{\labelimage[label=\textcolor{black}{(c)}]{\includegraphics[width=0.25\textwidth]{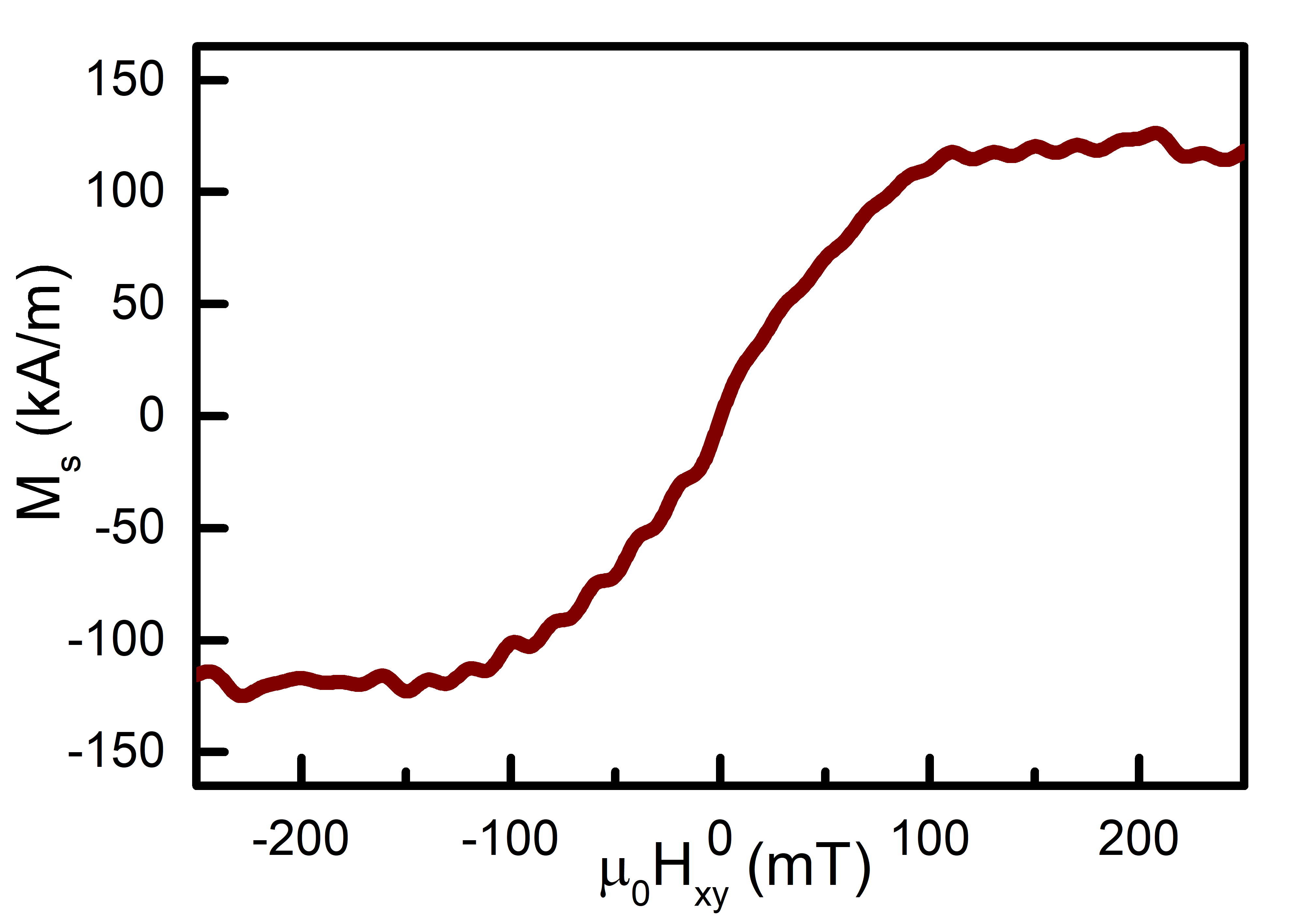}}\label{MvH15nm}}
\subfloat{\labelimage[label=\textcolor{black}{(d)}]{\includegraphics[width=0.25\textwidth]{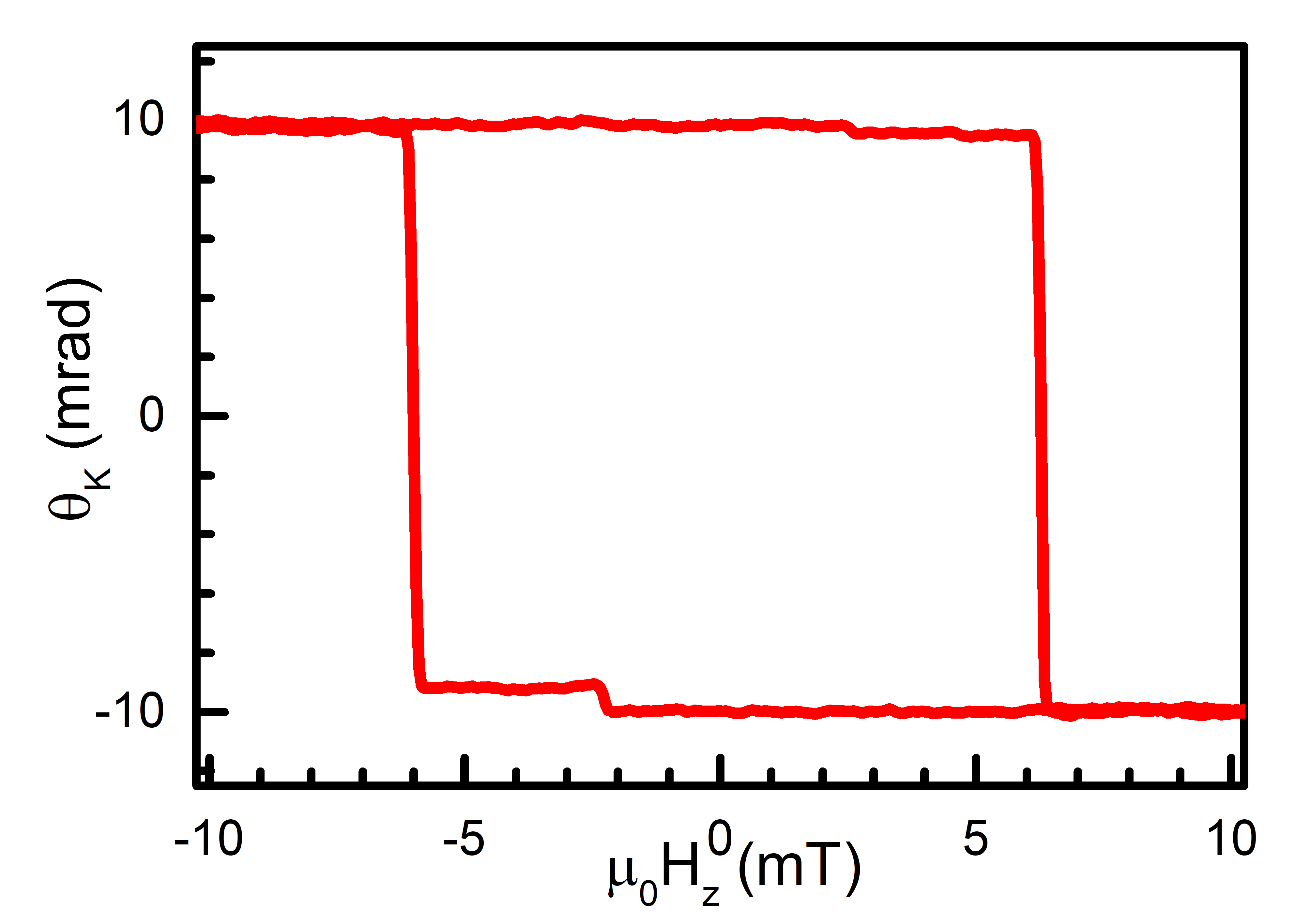}}\label{MOKE15nm}}
\hfill

\caption{\label{MvH}\textbf{Magnetic characterization of YIG grown on GSGG(111) substrate}. Magnetization vs. magnetic field in the in-plane direction (hard axis) of the film a) 10\,nm and c) 15\,nm YIG films. Polar MOKE hysteresis loop with magnetic field applied in the out-of-plane direction (easy axis), with a 3.0\,eV photon energy probe, for b) 10\,nm and d) 15\,nm YIG films.} 
\end{figure}

To assess the magnetic properties of the thin films, we measured the magnetization vs. external magnetic field (MvsH) hysteresis loops on the samples using a SQUID magnetometer. The magnetic field was applied in the in-plane direction. The obtained hysteresis loops display a slow rotation of the magnetization vector until it reaches saturation, corresponding to the magnetization being aligned along the hard-axis of the samples, demonstrating the perpendicular magnetic anisotropy (PMA) in the samples (Fig.~\ref{MvH10nm} and \ref{MvH15nm}). 
The measured moment of the samples is $\approx$ 100\,kA/m, which is in accordance to the reported values in literature \cite{Du2013}. We also performed polar magneto-optical Kerr effect (MOKE) experiments, using a probe wavelength of $\lambda = 400$\,nm. The $ \theta_K$ vs.~$H_z$ plots shown in Fig.~\ref{MOKE10nm} and \ref{MOKE15nm} furthermore corroborating the PMA in the YIG samples. It can be observed that the sign of the Kerr rotation angle for YIG is negative, as the angle of rotation is positive for negative external magnetic fields.

\subsection{\label{sec:PumpProbemeas}Pump-probe measurements}

For the time-resolved magneto-optic Kerr effect (TRMOKE) measurements, we used a pump-probe system built around a Ti:sapphire laser with a repetition rate of 75\,MHz. The amplitude of the probe beam ($\lambda = 400$\,nm) is modulated by means of a photo-elastic modulator (PEM) at a frequency of 120\,kHz, while a mechanical chopper modulates the amplitude of the pump beam ($\lambda = 800$\,nm) at a frequency of 5.95\,kHz. The pump power used was 5\,mW (unless specified otherwise), and the probe power used was 1\,mW. The spot size for both pump and probe beams are $w_o \approx 5\,\mu$m. A delay stage controls the arrival time of the pump beam pulses relative to the probe pulses. Both pump and probe beams are incident normal to the sample surface (i.e., $\theta = 0\degree$) on the top surface of the metal film. We control the polarization of the pump beam (linear polarization (LP), left-hand circular polarization (LCP), or right-hand circular polarization (RCP)) by means of a liquid crystal variable retarder (LCVR), carefully tuned to account for the added ellipticity of the reflecting surfaces in the experimental setup. We record the change in the Kerr rotation ($\Delta \theta_K$) of the probe beam induced by the pump pulse using a balanced photodetector. We make up for any added ellipticity in our measurement due to the reflecting surfaces in the detection line by means of a Babinet-Soleil compensator.

\section{\label{sec:Results}Results}

\subsection{\label{sec:LSSE}Heat-driven magnetization change in YIG/Au}

\begin{figure}[h]
\setlabel{pos=nw,fontsize=\normalsize} 
  \subfloat{\labelimage[label=\textcolor{black}{(a)}]{\includegraphics[width=0.40\textwidth]{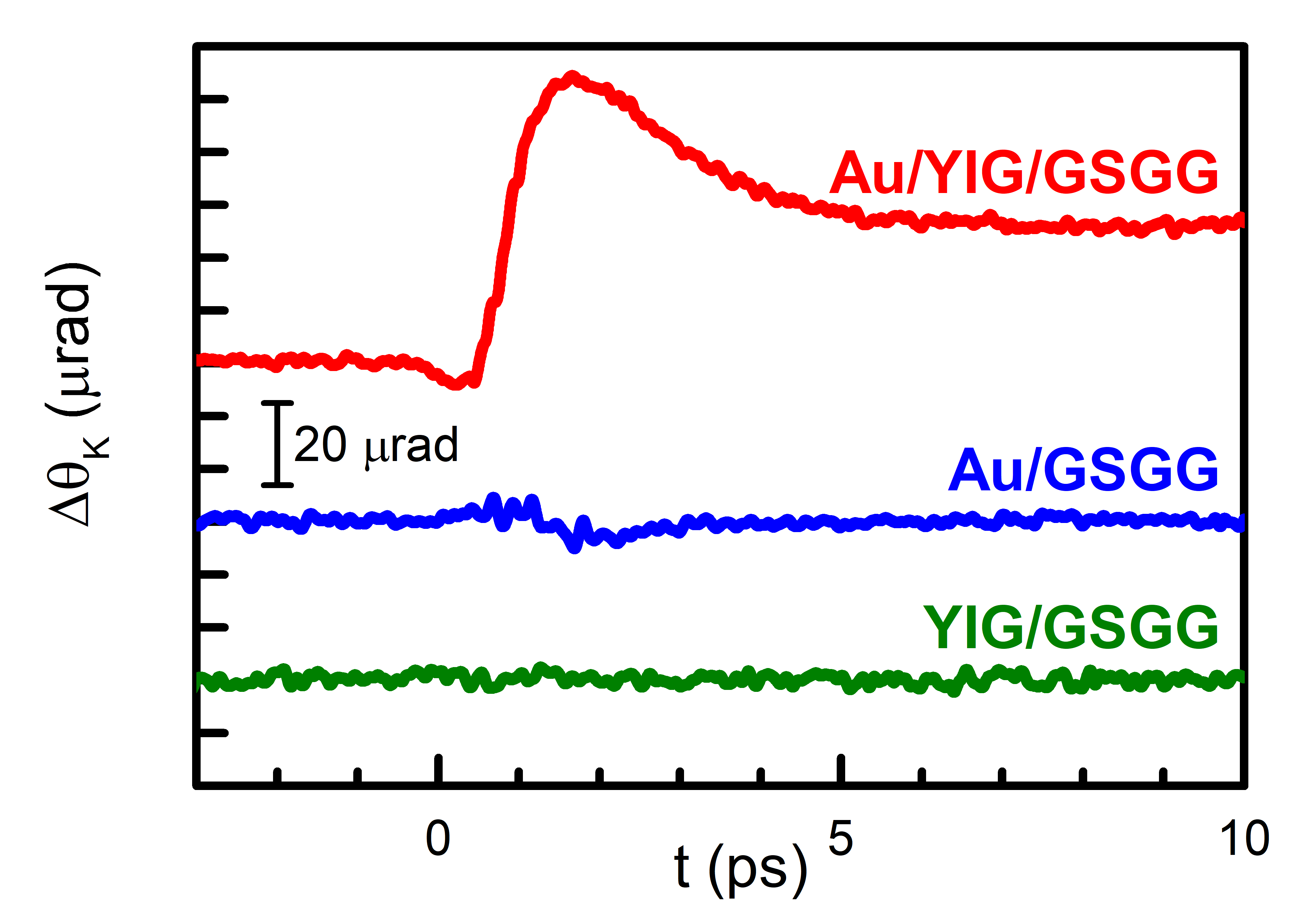}}\label{TRMOKES_all_mats}}
  \hfill
  \setlabel{pos=nw,fontsize=\normalsize}
  \subfloat{\labelimage[label=\textcolor{black}{(b)}]{\includegraphics[width=0.40\textwidth]{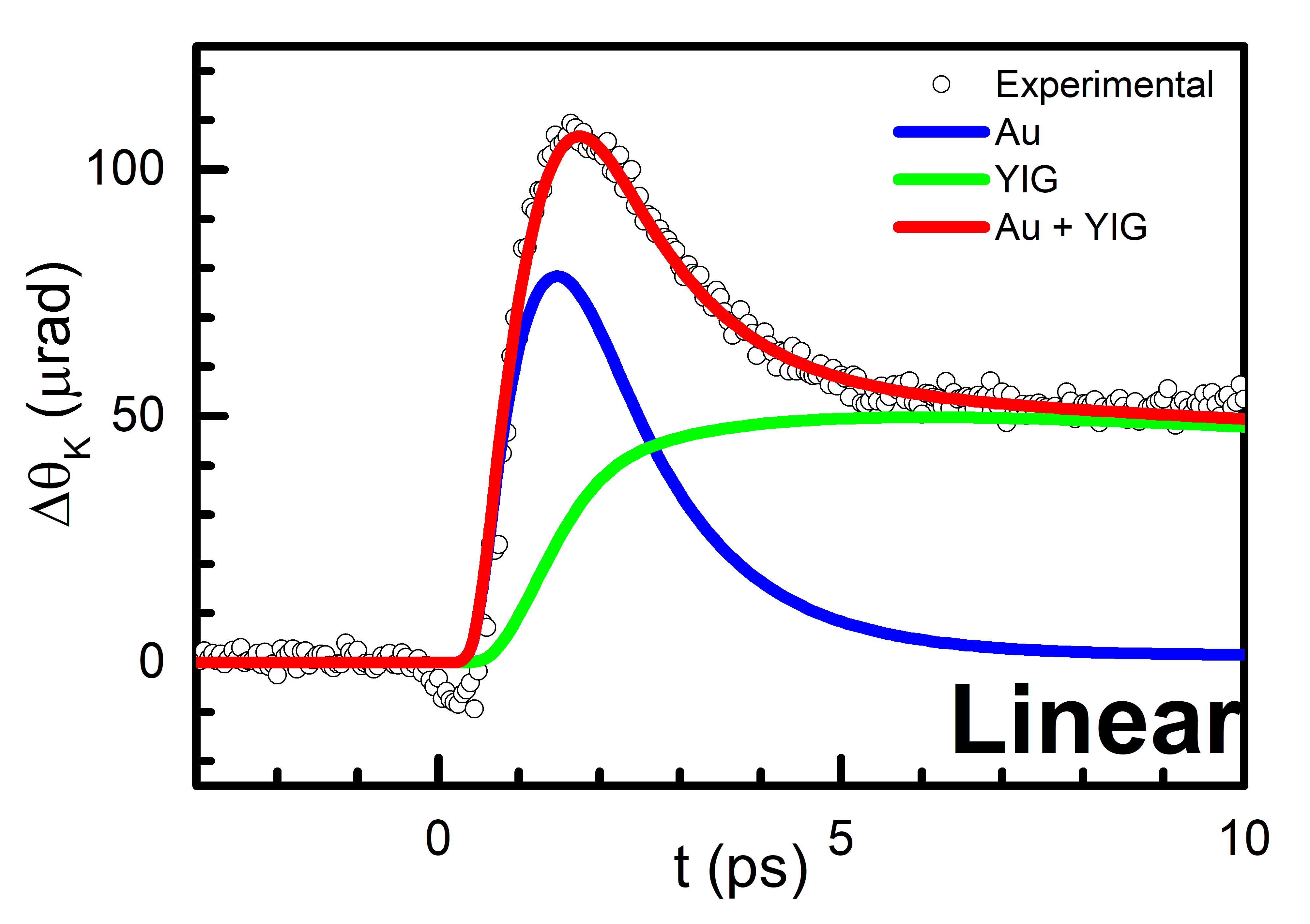}}\label{Simulation}}
  \hfill
  \setlabel{pos=nw,fontsize=\normalsize}
  \subfloat{\labelimage[label=\textcolor{black}{(c)}]{\includegraphics[width=0.40\textwidth]{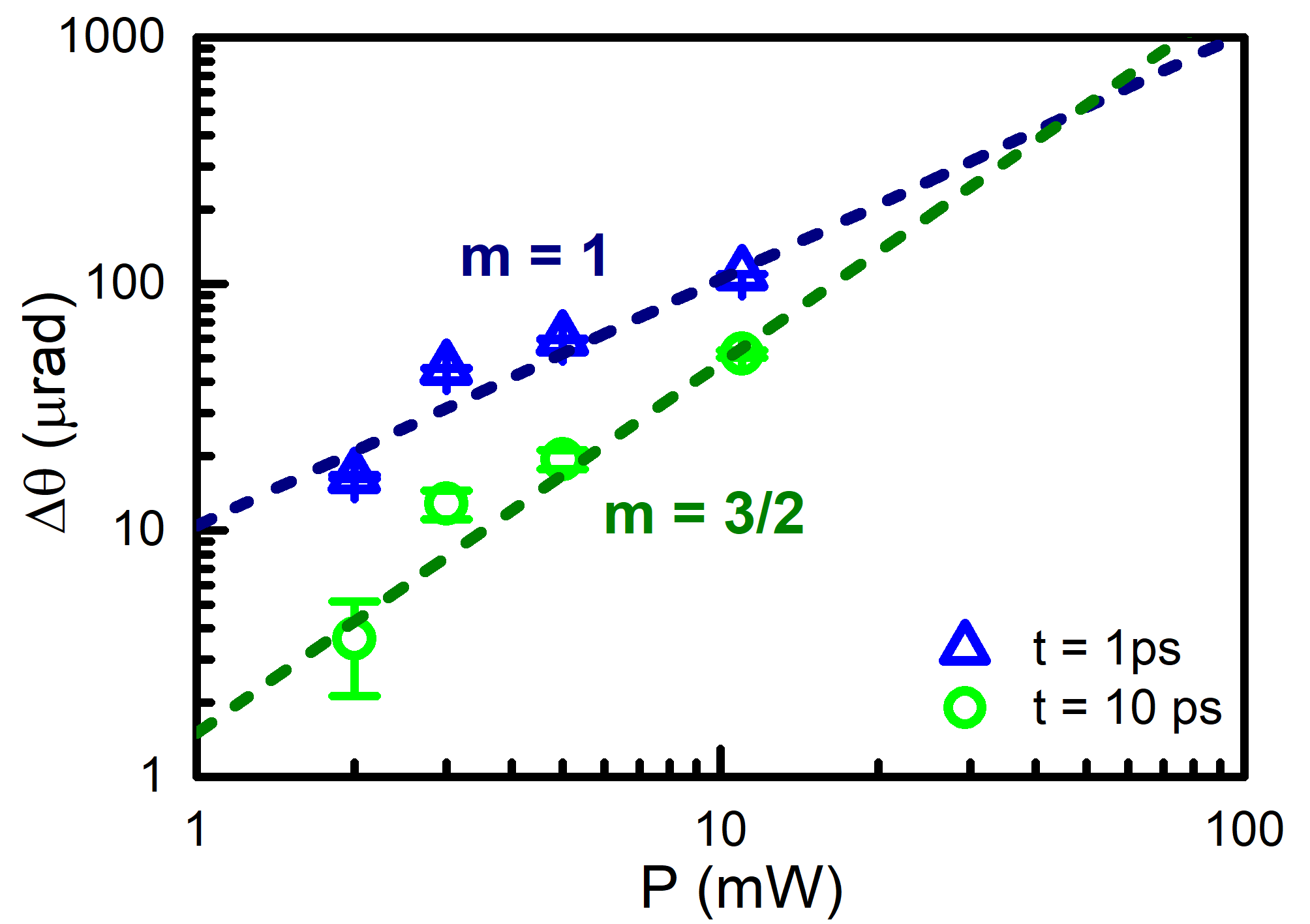}}\label{Powerdep}}
  \hfill
  \caption{\label{TRMOKE15nm}\textbf{Time-resolved MOKE for YIG(15nm)/Au(5nm).} a) Time-resolved MOKE for linear polarized pump beam for Au/YIG/GSGG, Au/GSGG and YIG/GSGG. b) Experimental data and simulation of the time evolution of the Kerr rotation ($\Delta \theta$) of the Au/YIG bilayer using the two-temperature and spin-diffusion models. c) Log-log plot of the Kerr rotation ($\Delta \theta$) vs. pump power P.}
\end{figure}

We first investigate the thermally-driven spin dynamics using a linearly-polarized (LP) pump pulse (Fig.~\ref{TRMOKE15nm}). Figure \ref{TRMOKES_all_mats} shows time delay scans of the change in Kerr rotation from equilibrium ($\Delta \theta_K$) for the Au/YIG/GSGG bilayer sample (red curve) and control samples with individual layers: Au/GSGG (blue curve) and YIG/GSGG (green curve).
An external magnetic field of $\mu_0 H_z$ = $\pm$10\,mT is applied to fully saturate the magnetization on the magnetic samples as seen in Fig.~\ref{MvH}). For the YIG/GSGG control sample, we observe no TRMOKE signal, as expected due to the negligible absorption of YIG at the 1.55\,eV photon energy of the pump beam. For the Au/GSGG control sample, we observe a very small signal that might be due to minor crosstalk with the thermoreflectance signal of gold. For the Au/YIG/GSGG bilayer, the TRMOKE results show a sudden increase in the Kerr rotation followed by a gradual reduction to a non-zero value.  Similar behavior has been observed previously in REIG/Au bilayers~\cite{Kimling2017,Ortiz2021} and can be understood within a two-temperature model (Fig.~\ref{Simulation}) as arising from thermally-induced spin transfer from YIG to Au via the longitudinal spin Seebeck effect (LSSE) (blue curve) and a change in the YIG magnetization (green curve) for a total simulated signal (red curve) that agrees with experiment (open circles). To further understand the origin of the measured signals, we perform pump-power dependence experiments (see Supplemental Information Fig.~S1~\cite{supp_material}) to evaluate the magnitudes of the signals at $t = 1$\,ps (maximum change in $\Delta \theta_K$) and $t = 10$\,ps (when $\Delta \theta_K$ stabilizes), as seen in Fig.~\ref{Powerdep}. 
The best fit yields a power law of $m=1.0$ at 1\,ps and $m=\frac{3}{2}$ at 10\,ps. The two-temperature model and pump-power dependence are further discussed in section \ref{sec:discussion}.

\subsection{\label{sec:stIFE}Angular momentum transfer due to IFE from Au to YIG}

\begin{figure}[h]
\setlabel{pos=nw,fontsize=\normalsize} 
   \subfloat{\labelimage[label=\textcolor{black}{(a)}]{\includegraphics[width=0.40\textwidth]{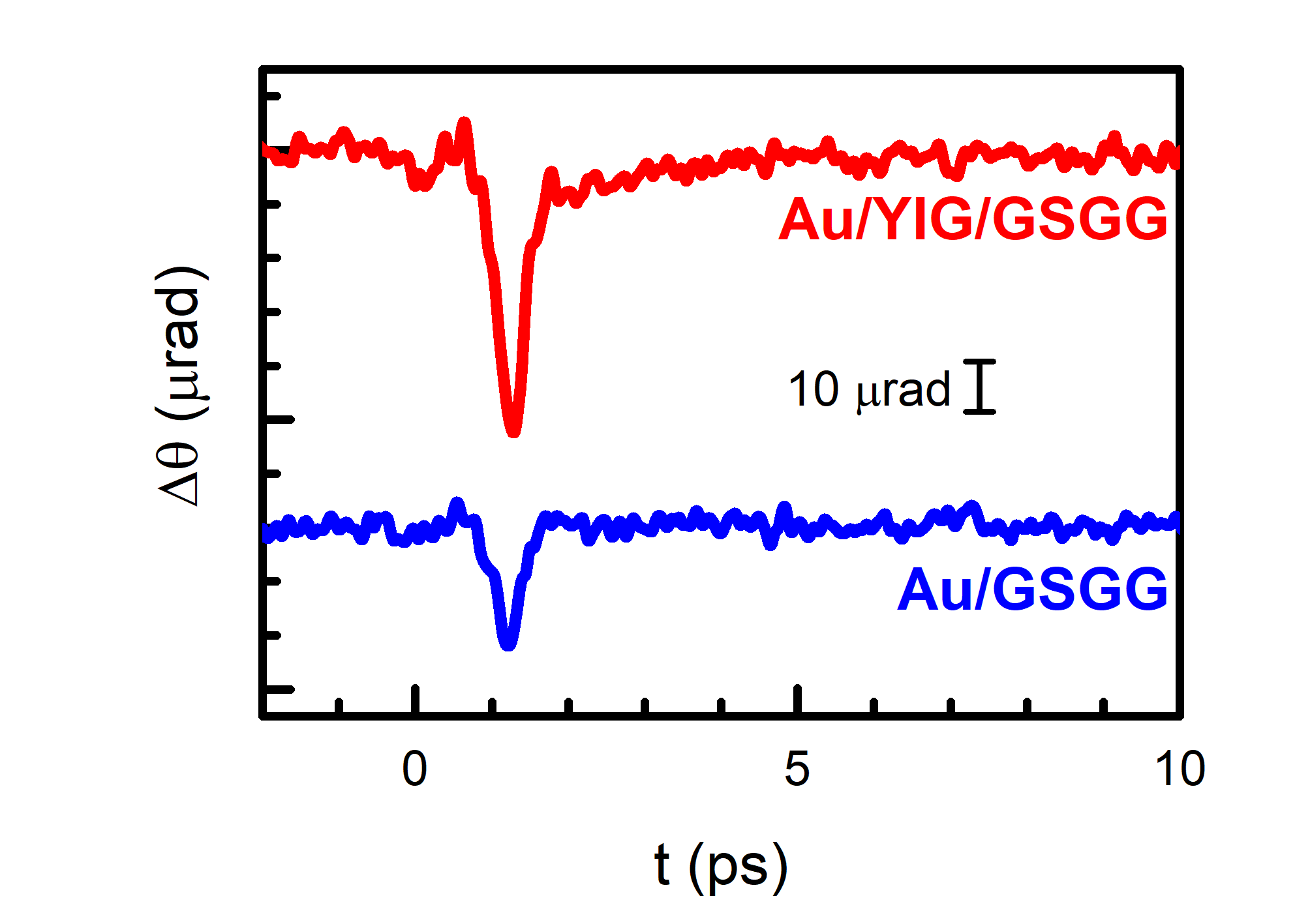}}\label{AuvsAuYIGall}}
  \hfill
  \setlabel{pos=nw,fontsize=\normalsize}
  \subfloat{\labelimage[label=\textcolor{black}{(b)}]{\includegraphics[width=0.40\textwidth]{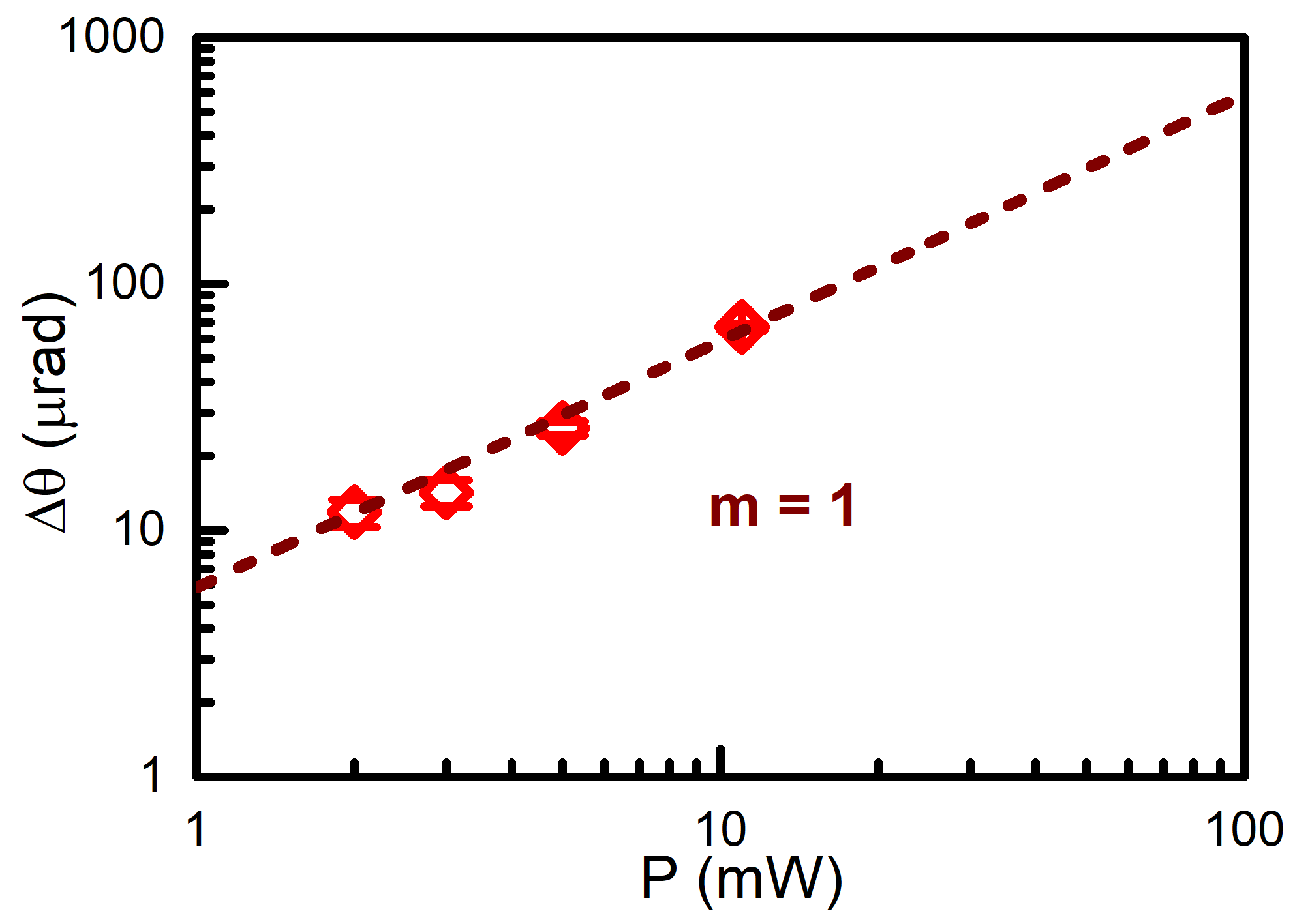}}\label{PowerdepIFE_10nm}}
  \hfill
  \caption{\label{AuvsAuYIG}\textbf{Inverse Faraday Effect on YIG/Au bilayer).} a) Zero-field IFE response for GSGG(111)-grown Au(5\,nm)/YIG(10\,nm), Au(5\,nm) and YIG(10\,nm) thin films. b) Log-log plot for the power dependence of maximum IFE response on YIG/Au bilayer.}
\end{figure}

We measure the TRMOKE responses to an RCP pump and an LCP pump, and subtract the two time-delay scans, in order to investigate the IFE on the samples. Figure \ref{AuvsAuYIGall} shows the zero-field IFE response of the Au(5nm)/YIG(10nm)/GSGG(111) bilayer sample, and compare with the Au(5nm)/GSGG(111) and YIG(10nm)/GSGG(111) reference samples. The measurement conditions are the same for all experiments. By analyzing the RCP - LCP signal of the experiments, we observe that: 1) the bare YIG sample does not show IFE response (green curve) which indicates that the intrinsic IFE of YIG under the current experimental conditions is negligible 2) the signal for the Au film (blue curve) indicates the presence of IFE-induced spin polarization in the Au film, and the observed temporal dependence is consistent with previous IFE experiments on non-magnetic transition metal films~\cite{Ortiz2023}, 3) the peak signal for Au/YIG sample (red curve) is 2.5 times stronger compared to the Au/GSGG sample. We measure the dependence of the peak value of IFE on the pump power (Supplemental Fig.~S2~\cite{supp_material}), which exhibits a power law of $m = 1.0$, as shown in Figure~\ref{PowerdepIFE_10nm}.

The previous observations on the Au/YIG experiments have revealed so far two contributions: 1) the spin injection due to the LSSE 2) the direct IFE signal arising from spin polarization in the Au layer. However, there is the possibility of a change in the magnetization of YIG induced by IFE in the Au layer. Here, the spin polarization generated by IFE in the Au layer transfers to the YIG layer and alters its magnetization. We denote this process as \enquote{spin-transfer IFE}.

\begin{figure}[h]
\setlabel{pos=nw,fontsize=\normalsize} 
  \subfloat{\labelimage[label=\textcolor{black}{(a)}]{\includegraphics[width=0.40\textwidth]{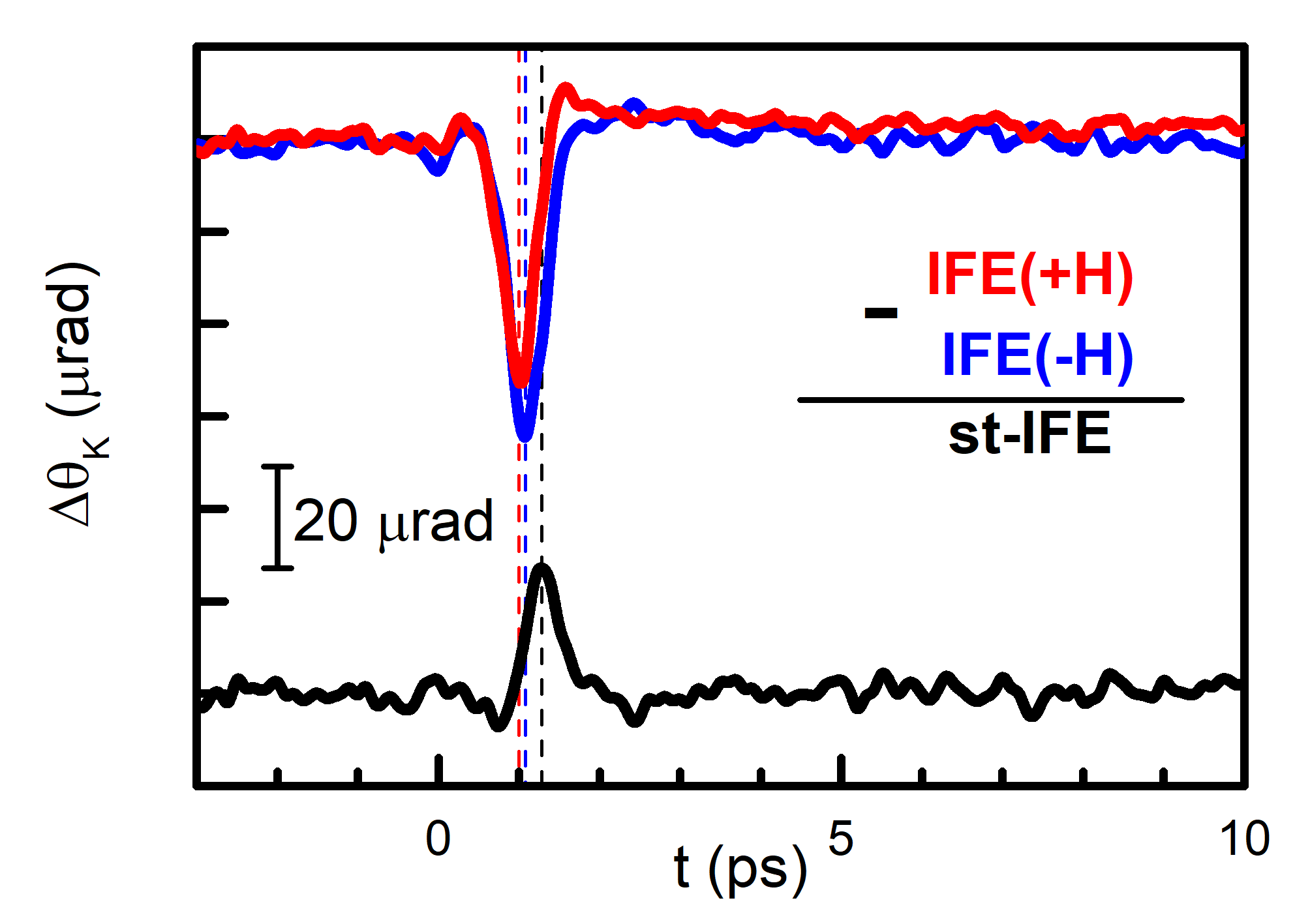}}\label{IFE_10nm_fields}}
  \hfill
  \setlabel{pos=nw,fontsize=\normalsize}
  \subfloat{\labelimage[label=\textcolor{black}{(b)}]{\includegraphics[width=0.40\textwidth]{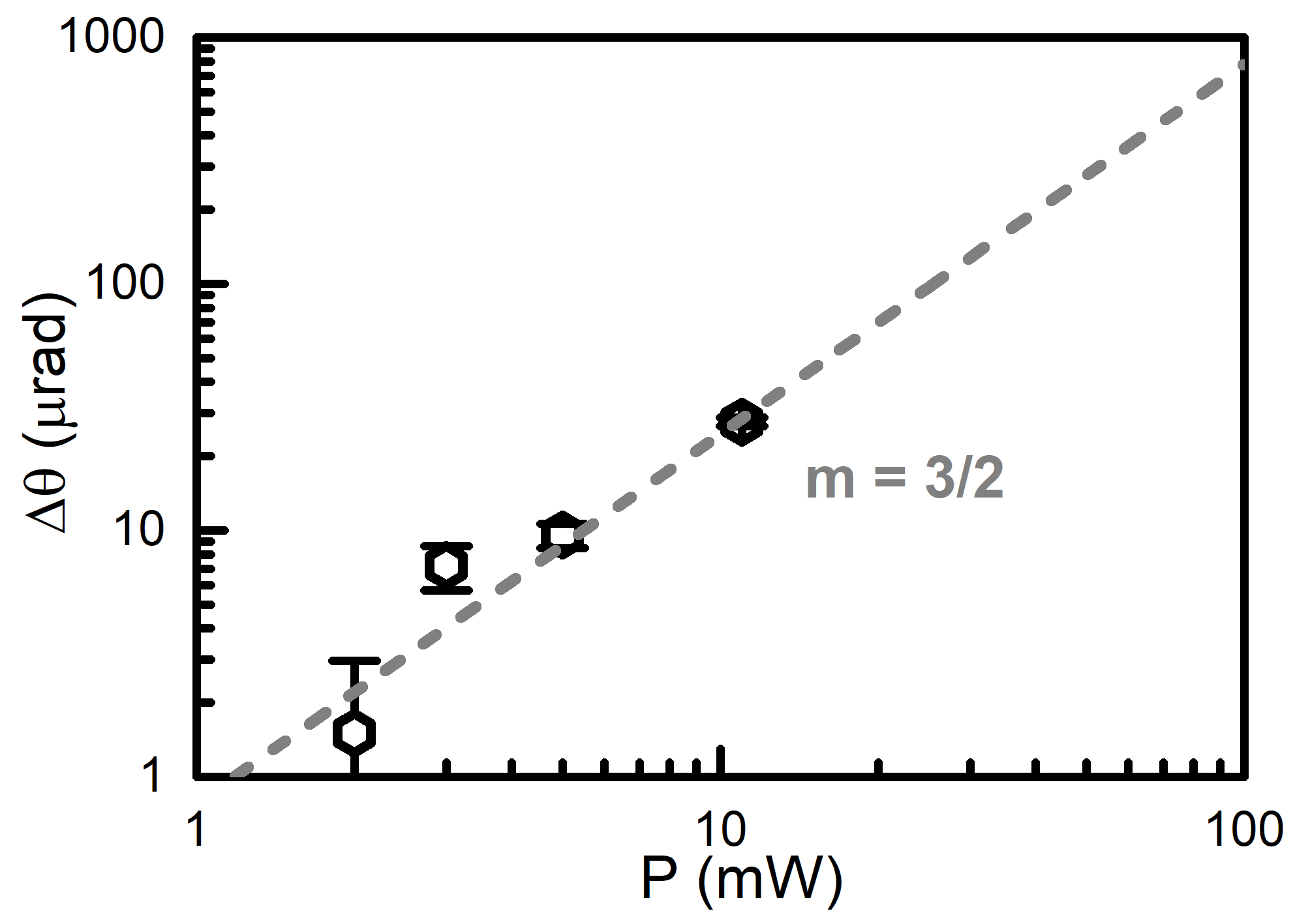}}\label{Powerdependence_stIFE_fit}}
  \hfill
  \caption{\label{TR-stIFE}\textbf{Spin-transfer inverse Faraday Effect on YIG/Au bilayer).} a) IFE response on Au/YIG bilayer with external magnetic field of $\pm$10\,mT, and the difference between both signals. This difference is referred as spin-transfer IFE. The dashed lines are a guide to the eye. b) Log-log plot for the power dependence of maximum st-IFE response on YIG/Au bilayer.}
\end{figure}

To isolate the spin-transfer IFE, we perform the IFE experiment in external magnetic fields of $\mu_0 H_z=\pm 10$\,mT, which is sufficient to magnetically saturate the YIG along the out-of-plane axis. Because the magnetization states along $+z$ and $-z$ are related by time-reversal, the IFE signals from YIG will be antisymmetric in $H_z$, while the IFE signals for Au are relatively insensitive to the external field because Au is diamagnetic and exhibits negligible magnetization in these low fields. Thus, subtracting the IFE signals from positive and negative $H_z$, denoted as IFE($\Delta H$), yields a signal that is dominated by the Kerr rotation signal from the YIG layer. Figure \ref{IFE_10nm_fields} shows the IFE signal for $+10$\,mT (red curve) and $-10$\,mT (blue curve), measured with an average pump power of $P = 11$\,mW. We can clearly observe a difference between the IFE signal measured at positive and negative applied external fields. We can disregard any cross-phase modulated (XPM) artifacts from the magnetic field as these effects are only dependent on light intensity~\cite{Lorenc2002} and independent of the external magnetic field. By subtracting the time-resolved signals with positive and negative fields, the remaining signal (black curve) shows a peak magnitude of $\approx 20\,\mu \text{rad}$ and a duration of $t \approx 2$\,ps. To understand the origin of this signal, we perform pump power dependence measurements of the spin-transfer IFE experiment (Supplemental Fig.~S3~\cite{supp_material}), and we plot the maximum amplitude of the spin-transfer IFE (Fig.~\ref{Powerdependence_stIFE_fit}). The analysis of these results is discussed further in section \ref{sec:discussion}.
 
To further understand the spin-transfer IFE signals, we proceed to perform magnetic field sweeps at different time delays for our time-resolved signals (we will refer to this type of experiment as time-resolved, field-dependent, or TRFD). 

\begin{figure*}
\setlabel{pos=nw,fontsize=\normalsize} 
    \subfloat{\labelimage[label=\textcolor{black}{(a)}]{\includegraphics[width=0.33\textwidth]{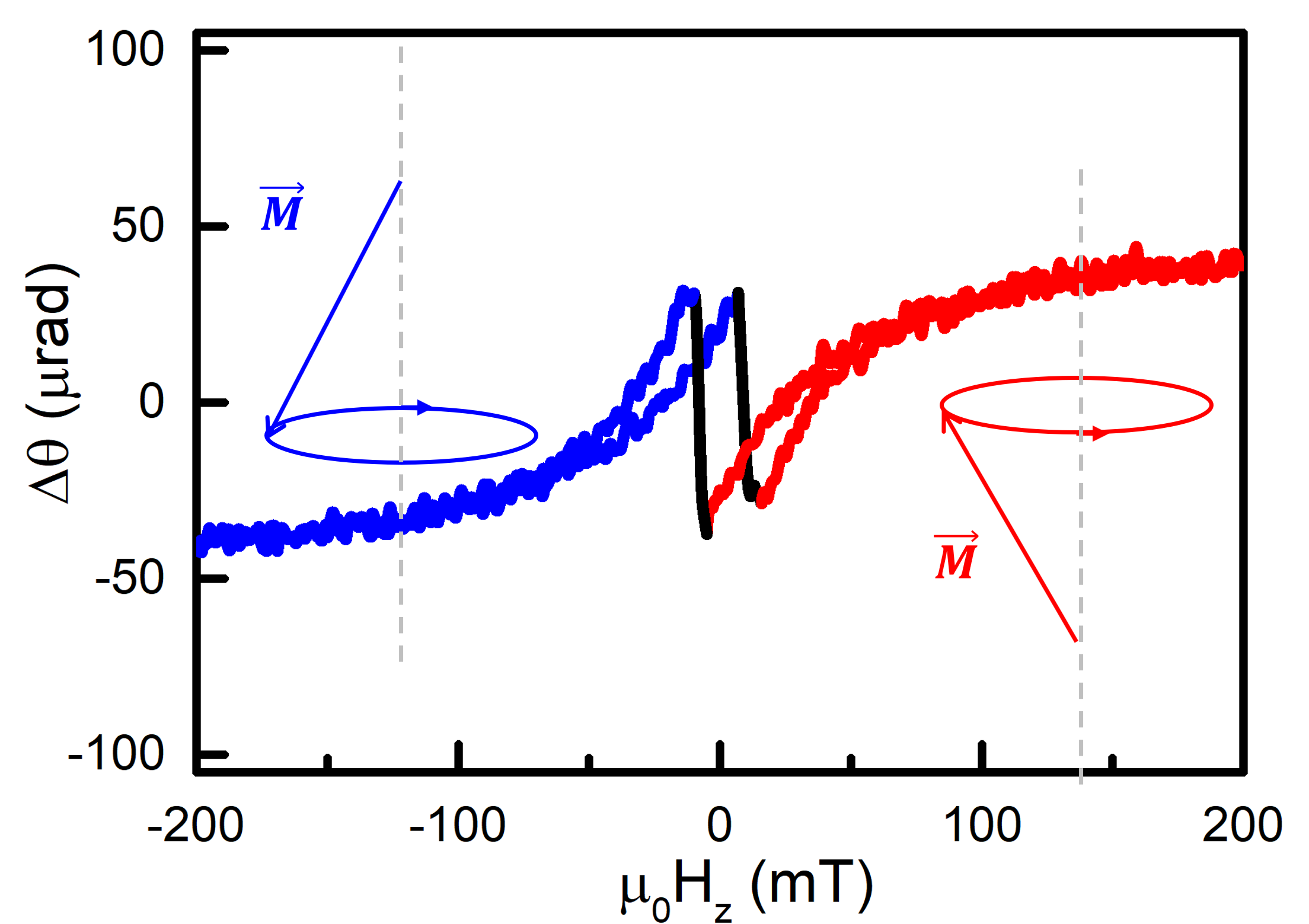}}\label{trfdMOKE}}


\subfloat{\labelimage[label=\textcolor{black}{(b)}]{\includegraphics[width=0.33\textwidth]{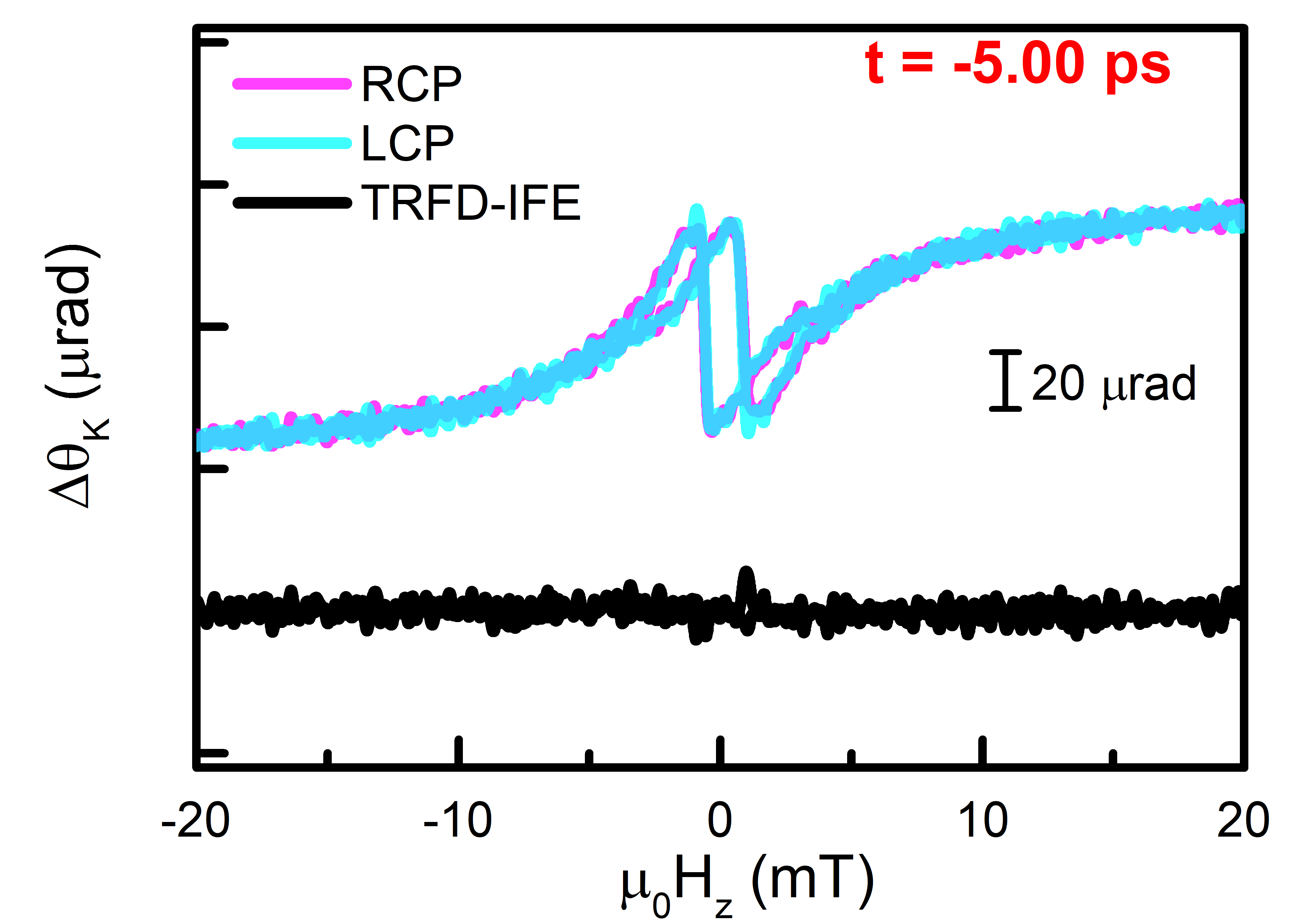}}\label{trfd_IFE_10nm_-3ps}
}
\subfloat{\labelimage[label=\textcolor{black}{(c)}]{\includegraphics[width=0.33\textwidth]{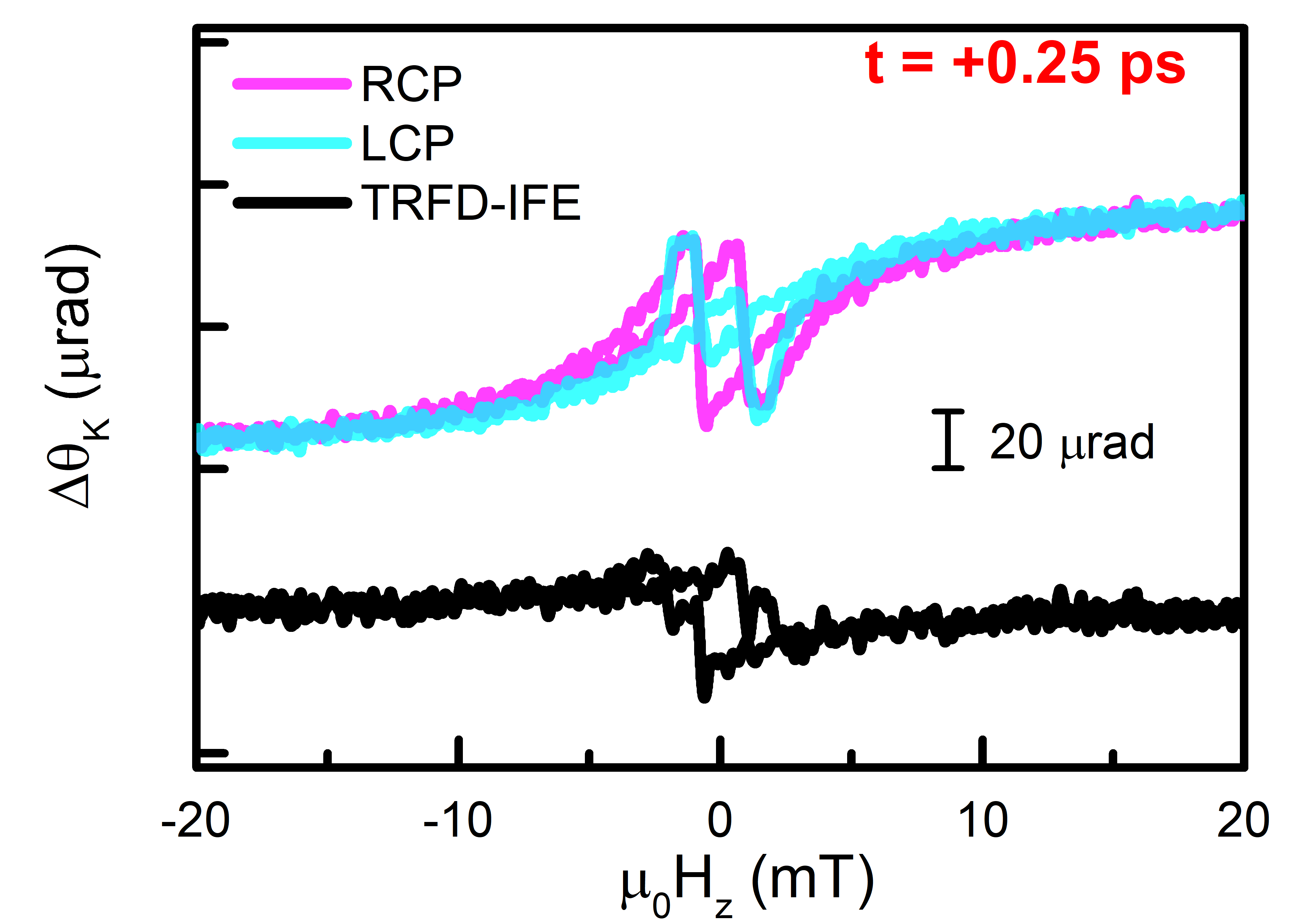}}\label{trfd_IFE_10nm_+0.25ps}
}
\subfloat{\labelimage[label=\textcolor{black}{(d)}]{\includegraphics[width=0.33\textwidth]{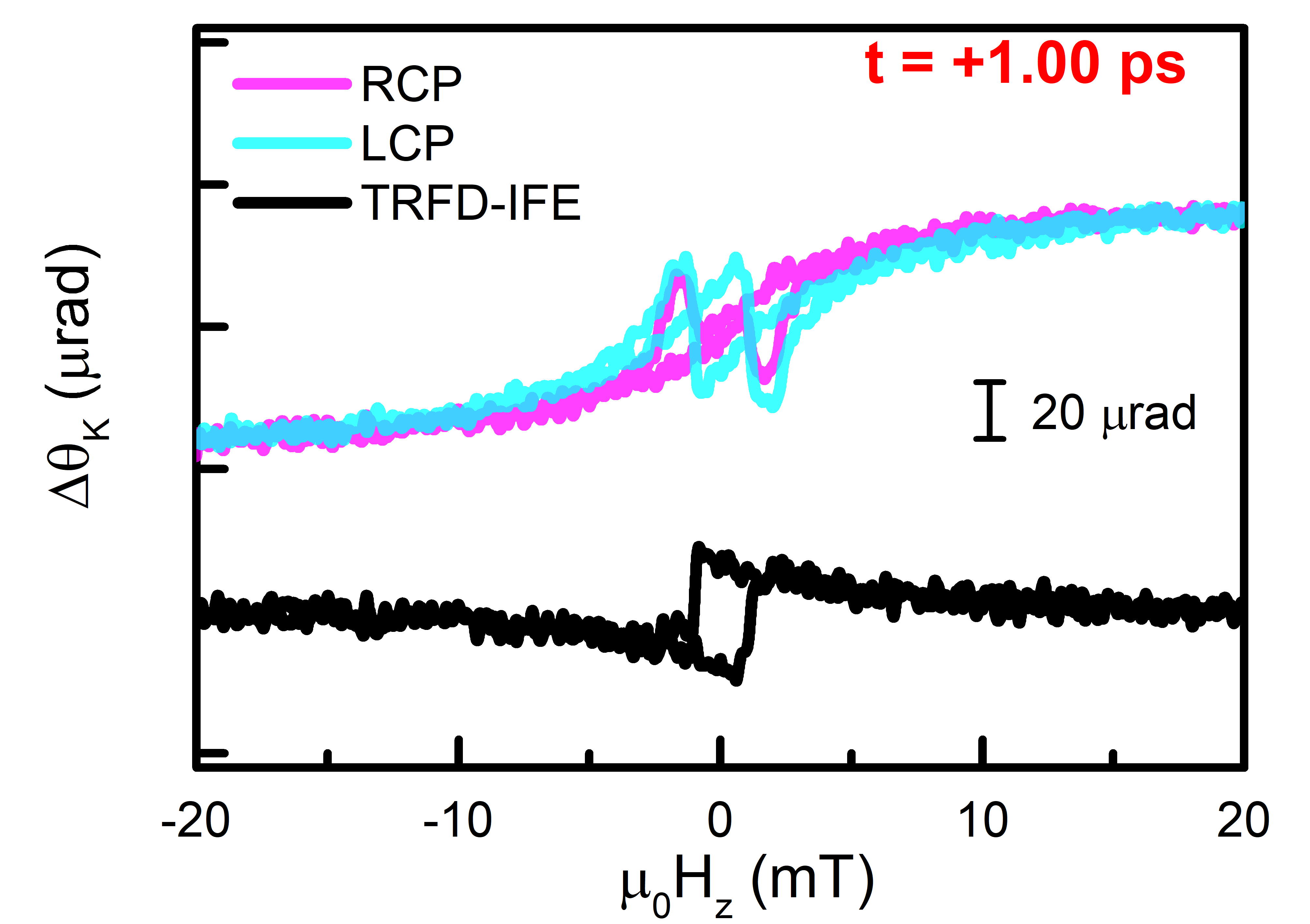}}\label{trfd_IFE_10nm_+1ps}
}


\subfloat{\labelimage[label=\textcolor{black}{(e)}]{\includegraphics[width=0.33\textwidth]{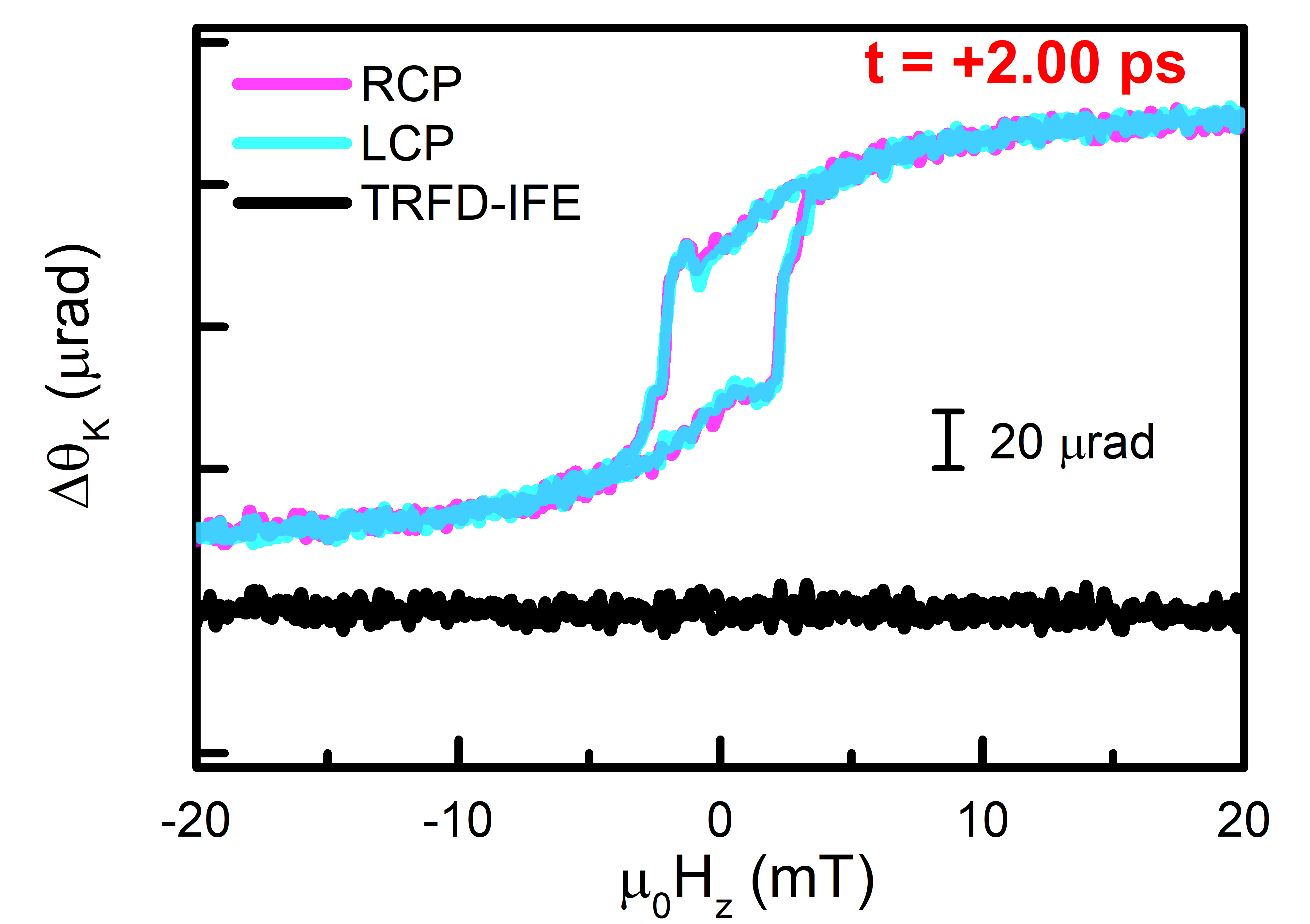}}\label{trfd_IFE_10nm_+2ps}
}
\subfloat{\labelimage[label=\textcolor{black}{(f)}]{\includegraphics[width=0.33\textwidth]{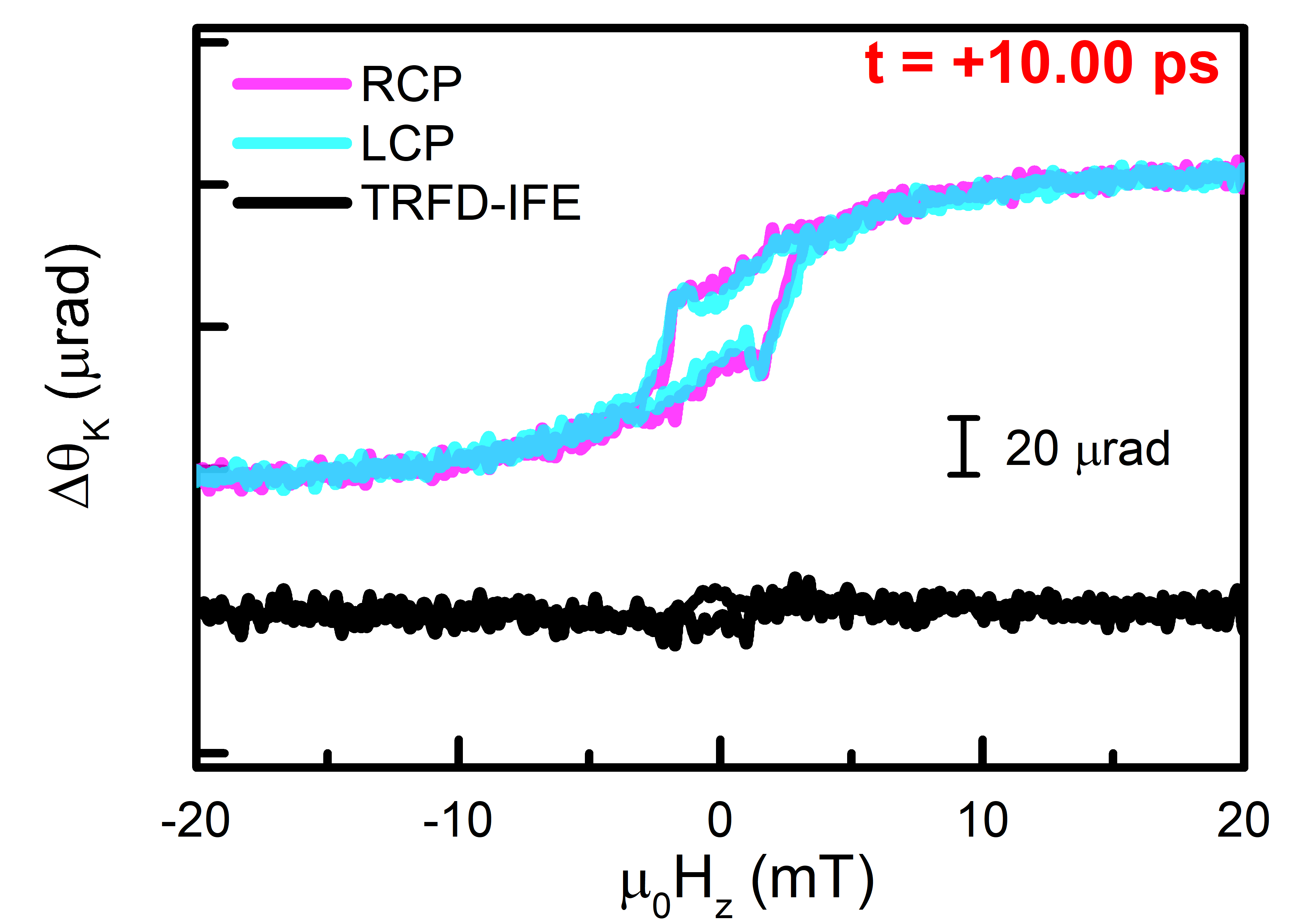}}\label{trfd_IFE_10nm_+10ps}
}
\hfill
\subfloat{\labelimage[label=\textcolor{black}{(g)}]{\includegraphics[width=0.33\textwidth]{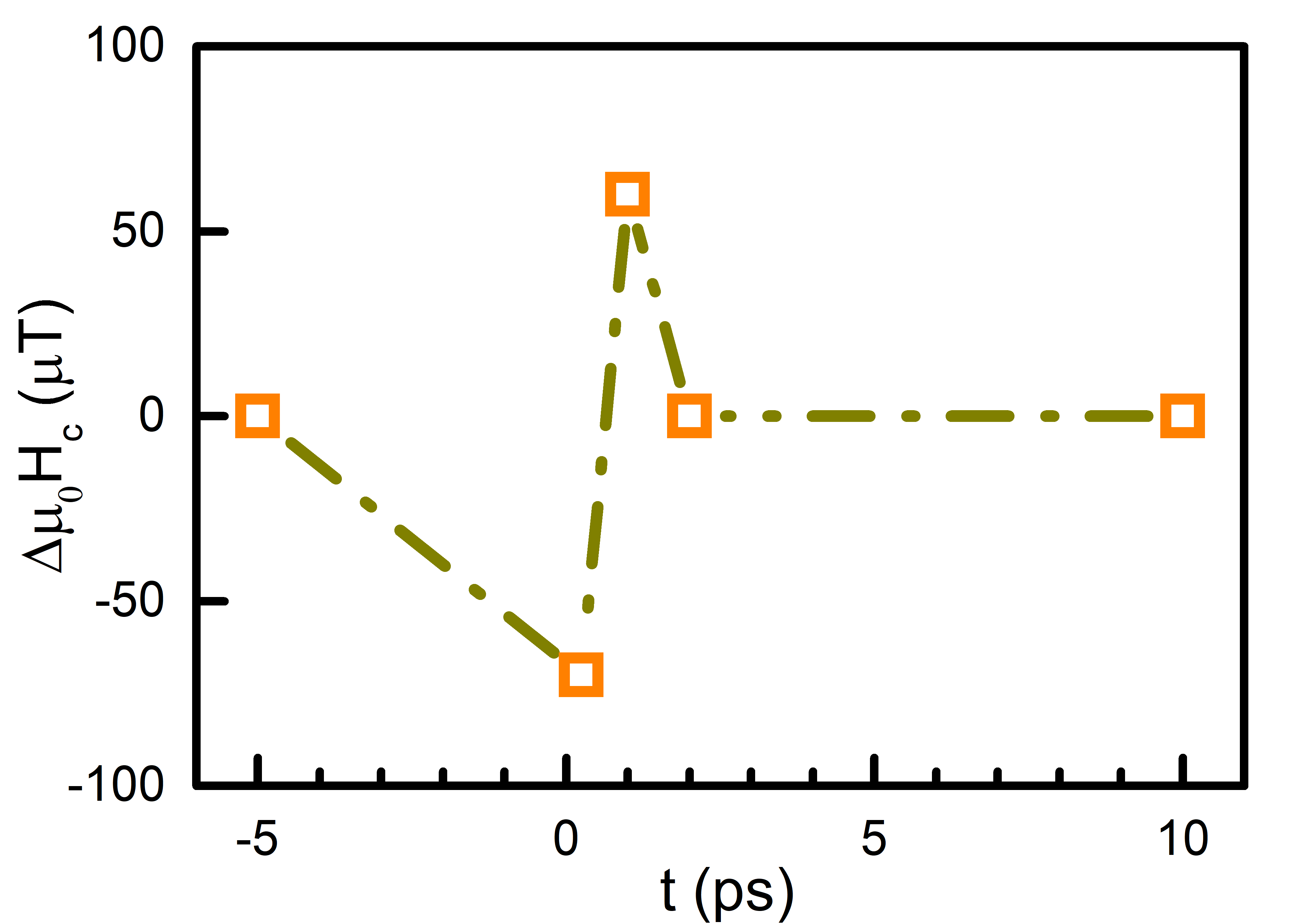}}\label{Hc10nm}
}
\subfloat{\labelimage[label=\textcolor{black}{(h)}]{\includegraphics[width=0.33\textwidth]{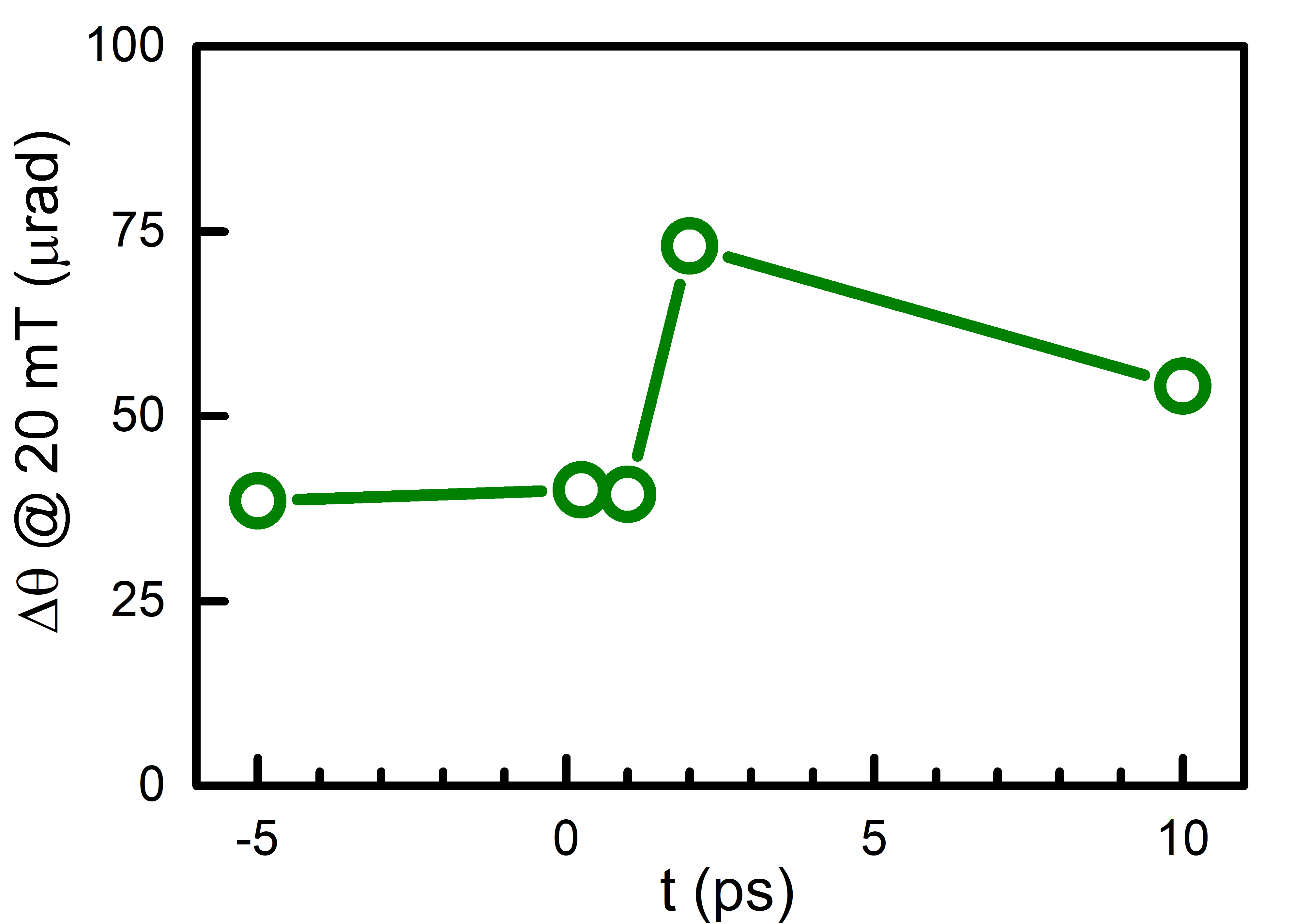}}\label{dtheta10nm}
}

  
\caption{\textbf{Spin-transfer Inverse Faraday Effect on Au(5\,nm)/YIG(10\,nm) bilayer.} a) Time-resolved, field dependent magneto-optical Kerr effect (TRFD MOKE) hysteresis loop, displaying the magnon behavior in YIG at a given time $t$ as the magnetic field is swept. In this example, the TRFD MOKE corresponds to RCP pump excitation at t = +0.25 ps. b-f) TRFD MOKE hysteresis loops for right and left circular polarized pump beam at different pump/probe delay times. The black line represents the IFE-related signal (RCP - LCP), g) difference of coercive fields $H_c$ at different pump/probe time delays, h) maximum amplitude of the TRFD MOKE hysteresis loops with respect to the pump/probe time delay.}

\label{TRFD10nm}
\end{figure*}

For a given pump-probe time delay $t$, the magnetic field dependent Kerr rotation hysteresis loop $\left( \Delta \theta_K \;\text{vs.}\; H_z\right)$ can be seen in Fig.~\ref{trfdMOKE}. The measured loop can be interpreted as the ultrafast magnetic state of the YIG/Au bilayer, similar to what was observed in a similar experiment on a Ni thin film \cite{Beaurepaire1996}. Each red and blue section of the hysteresis loops corresponds to magnon precession aligned to the external magnetic field, while the black sections correspond to the magnetization flipping of YIG.  

For the TRFD measurements, we select specific time delays based on the time-resolved LSSE data (Fig.~\ref{Simulation}) and time-resolved IFE data (Fig.~\ref{IFE_10nm_fields}):
we probe the sample at $t = -5.00$\,ps (probe arrives earlier than pump) (Fig.~\ref{trfd_IFE_10nm_-3ps}), $t = +0.25$\,ps (IFE signal starts raising) (Fig.~\ref{trfd_IFE_10nm_+0.25ps}), $t = +1.00$\,ps (maximum IFE signal) (Fig.~\ref{trfd_IFE_10nm_+1ps}), $t = +2.00$\,ps (IFE signal has decayed, while IFE($\Delta$H) and LSSE signals are maximum) (Fig.~\ref{trfd_IFE_10nm_+2ps}), and $t = +10.00$\,ps (bilayer has reached thermal equilibrium and YIG is cooling down) (Fig.~\ref{trfd_IFE_10nm_+10ps}). We perform the experiments using RCP and LCP pump light (magenta and cyan curves, respectively), and then evaluate the difference between both signals (black curve in Fig.~\ref{trfd_IFE_10nm_-3ps}-\ref{trfd_IFE_10nm_+10ps}). In this way, we are able to distinguish the thermal effect contributions (which are independent of the polarization of the pump) and the IFE-related contributions. 

To further evaluate the influence of the chirality of the CP pump, we get the difference between the coercive fields $H_c$ (Fig.~\ref{Hc10nm}). We also evaluate the time evolution of the maximum amplitudes of the TRFD MOKE loops (Fig.~\ref{dtheta10nm}). These data are discussed in detail in the next section.

\section{\label{sec:discussion}Discussion}

For the LSSE experiment using LP light (Fig.~\ref{TRMOKES_all_mats}), we model the change in temperature in the YIG/Au bilayer using the two-temperature model, using the thermal parameters in the Table SI of the Supplemental Information~\cite{supp_material}. This way, we estimate the change the Kerr rotation due to the spin injection from the YIG into the Au layer from the longitudinal spin-Seebeck effect (LSSE) and the change in magnetization of YIG (Fig.~\ref{Simulation}), assuming that the phonon-magnon thermalization time between 0.1 - 1 ps~\cite{Cornelissen2016}. From the simulated signal, we observe that the electrons and phonons in the Au layer have thermalized after 5 ps~\cite{Tas1994}, with the Au spins decaying due to spin-flip scattering. The remaining signal comes from the thermal demagnetization of the YIG layer (\cite{Kimling2017, Ortiz2021}) by the energy transfer to YIG magnons due to phonon-magnon coupling of YIG. By using the time-dependent spin-diffusion equation and the net Kerr angle in Au due to spin accumulation for $\lambda_{probe}$ = 400 nm~\cite{Ortiz2022}, we obtain the fitted parameters from the spin-diffusion model shown in Table SII~\cite{supp_material}. These values show that our calculated YIG/Au electron-magnon conductance $G_{e-m} = 1.29 \, \frac{MW}{m^2 K} $ and spin-relaxation time $\tau_{s} = 1.27$\, ps are consistent with previous studies~\cite{Kimling2017,Ortiz2021}.

To determine the effects of the pump polarization in the LSSE experiment, we analyzed the results of the experiments using CP pump light. The fitted spin-diffusion parameters can be found in Table SII~\cite{supp_material}. We found that the fitted $G_{e-m}$ values for all pump polarization states have similar values, which indicates that the spin-injection from the YIG into the Au layers is unaffected by the polarization state of the pump excitation. We also found that there is a difference of $\tau_s$ among the different polarizations about $\approx$ 100 fs. This difference in spin-relaxation time cannot be explained by the current spin dynamic models.

From the pump power dependence experiments, the $\Delta \theta_K$ vs $P_{pump}$ for the LP case shows for that for $t = 2$\,ps,  a linear fit of the log-log plot yields a slope of $m = 1.0$ (Fig.~\ref{Powerdep}), denoting the linear relation between the injected power and the spin injected into the gold, which has been observed in previous experiments where the transport of thermally-driven magnons is present~\cite{Adachi2011, Schreier2013, Guo2016, Jiang2016}. For $t = 10$\,ps, the log-log plot fits to a line indicating a power law of $\frac{3}{2}$. This is an indication that the observed signal after 5 ps comes from the YIG magnons only, as the electrons and phonons in the Au layer have thermalized by then, with the YIG's phonon-magnon interaction being the sole responsible of the signal; in other words, $m=\frac{3}{2}$ arises from the thermal response of YIG, as $M_\textrm{YIG} \sim ( \, 1-\frac{T}{T_C}) \,^{\frac{3}{2}}$ follows Bloch's spontaneous magnetization law~\cite{Bloch1930}. For the IFE experiment with zero applied field, the pump power dependence reveals a linear behavior with $m = 1.0$ (Fig.~\ref{PowerdepIFE_10nm}), in accordance to the semiclassical theory of IFE~\cite{Hertel2006}. However, for the spin-transfer IFE experiment, the pump power dependence shows that the data fits to a $m = \frac{3}{2}$ power law (Fig.~\ref{Powerdependence_stIFE_fit}). As the spin-transfer IFE experiment is performed with opposite magnetic fields, the obtained signal cannot be associated with the thermal spin-injection from YIG into Au due to the LSSE. 
In addition, in Fig.~\ref{IFE_10nm_fields}, the maximum amplitude of the spin-transfer IFE signal (peak) is shifted from the maximum amplitude (dip) of the IFE signal, indicating that the spin-transfer IFE signal is not a measurement artifact of the IFE signal from the Au layer.
Finally, the generation of magnetic signals by direct absorption in YIG is ruled out by the absence of TRMOKE signal in YIG/GSGG (Figs.~\ref{TRMOKES_all_mats} and \ref{AuvsAuYIGall}).
Thus, the evidence supports a mechanism where spin polarization generated by IFE in the Au layer transfers angular momentum to the YIG layer via a spin-transfer process.



The TRFD MOKE hysteresis loops shown in Fig.~\ref{TRFD10nm} can be viewed as a snapshot of magnons in the YIG/Au system at the time-delay the experiment is performed~\cite{Beaurepaire1996}. For $t<0$\,ps (i.e., before the CP pump beam excites the sample), the TRFD MOKE shows a hysteresis loop shown in Fig.~\ref{trfd_IFE_10nm_-3ps}. This signal proceeds from the magnons in YIG, which are a consequence of the thermal excitations in the previous pulse ($T_{pulse} = 13.3$\,ns). As we discussed previously, we do not expect contributions from LSSE because the Au phonons/electrons have reached thermal equilibrium by then. We confirm the thermal nature of this signal by getting the difference between the RCP and LCP pump TRFD response (we will refer to this RCP-LCP as the TRFD-IFE), which leads to no signal above the noise level.

At $t = +0.25$ ps (Fig.~\ref{trfd_IFE_10nm_+0.25ps}), the CP pump pulse excitation starts transferring angular momentum to the electrons in gold. At the same time, it causes an increase in the temperature of the Au layer, causing its electron temperature to increase. Because of the thermal gradient at the Au/YIG interface, there is a spin injection due to LSSE as discussed previously for the LP pump experiment. In the TRFD MOKE hysteresis loops, the thermal-only related contributions (LSSE spins in Au) should be the same for both LCP and RCP, therefore, the TRFD-IFE should reflect the change in the magnons of YIG due to the circular polarized excitation of the pump. At $t = +1$\,ps (Fig.~\ref{trfd_IFE_10nm_+1ps}), where the CP pump pulse is at its peak magnetic response, the TRFD-IFE shows a change of sign with respect to the TRFD-IFE at $t = +0.25$\,ps. The reason for this change of sign remains unclear at this time. A possibility is that the magnetic sites in YIG (tetrahedral and octahedral sites that are antiferromagnetically aligned) have different temporal responses to the spin-transfer IFE; however, the corroboration of this conjecture requires further theoretical and experimental studies. Finally, at $t > +2$\,ps (Fig.~\ref{trfd_IFE_10nm_+2ps} and \ref{trfd_IFE_10nm_+10ps}), the TRFD-IFE response goes back to zero, indicating that the IFE-related response of YIG has ended. 

 The evolution of the hysteresis loops of the TRFD MOKE experiments can be further analyzed by the clear change that is observed where the difference between the levels at $+20$\,mT and $-20$\,mT change (Fig.~\ref{dtheta10nm}). Compared with the spin-injection experiment (Fig \ref{Simulation}), we conclude that the amplitude of the TRFD MOKE experiments corresponds to both YIG magnons and the spins present in the Au layer. However, when comparing the coercive fields in the TRFD MOKE hysteresis loops, we observe that there is a clear difference between the LCP and RCP values for $t = +0.25$\,ps and $t = +1.00$\,ps (Fig.~\ref{Hc10nm}); moreover, the difference reverses sign between both cases. We observed a similar behavior for the YIG(15\,nm)/GSGG(111) sample (Supplemental Fig.~S4~\cite{supp_material}), with a similar evolution in the coercive fields for the TRFD experiment. Because gold lacks magnetic structure (i.e., domains), these changes in the hysteresis loops prove the modification of the magnetism of YIG. Because the spin accumulation (from both LSSE and IFE) in Au does not change sign during the time interval where the phenomenon occurs, we can disregard magnetic domain pinning in YIG due to the transient magnetism in the adjacent Au layer. 

 The results obtained in this work reveal that the spin-transfer IFE is capable of modifying the magnetization of the YIG layer by means of ultrafast excitations. For comparison, Hansteen \textit{et al.}~\cite{Hansteen2006} showed by time-resolved Faraday effect experiments that CP laser pulses with fluence $\sim$ 640 $J/m^2$ can induce oscillations in micrometer-thick doped YIG. In contrast, in our spin-transfer IFE experiment, we observe by means of TRMOKE measurements a modification in the magnetization of nanometer-thick undoped YIG with pump fluence as low as 6.8 $J/m^2$, although no oscillatory response was achieved in our case. These results invite further investigation by doing experiments with higher pump power, different metallic coatings, and/or different rare-earth iron garnets.
 
\section{\label{sec:conclusion}Conclusion}
 
Our experiments show the spin-transfer inverse Faraday effect (st-IFE) in  the Au/YIG bilayer, which consists on the transfer of spins from a metallic transducer (Au) to a ferrimagnetic insulator (YIG). These results show modification in the magnetic properties via magnonic excitations of YIG using lower excitation power compared to previous experiments. These results open a path for the use of iron garnet/metal bilayers for its implementation in infrared optospintronic technologies.  Moreover, insights into how interface quality, layer thickness, and optical parameters govern IFE strength will inform the design of low-power, high-speed memory and logic elements. Ultimately, a deeper understanding of spin-transfer IFE in metallic/ferromagnetic-insulator bilayers could unlock new regimes of coherent spin control, bridging photonics and magnonics for next-generation information technologies.

\section*{\label{sec:acknowledgments}Acknowledgments}
We would like to thank Prof. P. Oppeneer for fruitful discussion. This work was supported by the Center of Emergent Materials, an NSF MRSEC, under award number DMR-2011876. D.R. acknowledges partial support by the National Science Foundation under award number DMR-2225646 (YIG film growth).
The authors declare no conflict of interest.

\begin{table}[]
     \label{LSSE_fitted_parameters}
\end{table}
\bibliography{apssamp}

\end{document}